\documentclass[aps,pre,twocolumn,showpacs,floatfix]{revtex4}
\usepackage{latexsym,hyperref,color,amssymb,graphicx}

\newcommand{\thalf}{{\textstyle{\frac{1}{2}}}}

\newcommand{\beq}{\begin{equation}}
\newcommand{\eeq}{\end{equation}}
\newcommand{\bea}{\begin{eqnarray}}
\newcommand{\eea}{\end{eqnarray}}
\newcommand{\bean}{\begin{eqnarray*}}
\newcommand{\eean}{\end{eqnarray*}}
\newcommand{\bei}{\begin{itemize}}
\newcommand{\eei}{\end{itemize}}
\newcommand{\ben}{\begin{enumeration}}
\newcommand{\een}{\end{enumeration}}
\newcommand{\nn}{\nonumber}
\begin{document}

\title{Helices in Biomolecules}

\author{Kevin Cahill}
\email{cahill@unm.edu}
\affiliation{Department of Physics and Astronomy, 
University of New Mexico, Albuquerque, NM 87131}
\date{\today}
\begin{abstract}
Identical objects, regularly assembled, 
form a helix, which is the 
principal motif of
nucleic acids, 
proteins, and viral capsids.
\end{abstract}

\pacs{87.15.By,87.14.Ee,87.14.Gg,82.35.Pq}

\maketitle
\section{Helices in Biological Molecules}
Double-stranded DNA is 
a double helix~\cite{Watson1953}\@.
The principal secondary structures in proteins
are \(\alpha\)-helices~\cite{Pauling1951a} and 
\(\beta\)-sheets~\cite{Pauling1953a,Pauling1953b}, 
which are sheets of helices.
The fibrous protein \(\alpha\)-keratin
is a double \(\alpha\)-helix;
collagen is a triple helix.
The cytoskeletal filaments
--- actin filaments, microtubules, and intermediate
filaments --- are helical assemblies of subunits.
Helices occur in the
capsids of viruses.
\par
Helices are important and ubiquitous in biology
because identical objects, regularly assembled, 
form a helix.
This theorem --- that a regular assembly of
identical objects is a helix --- has been known 
in the biological community
since the work
of Pauling~\cite{Pauling1951a,Pauling1953a,Pauling1953b}
but is less familiar to physicists. 
Although it can be derived from the differential geometry of
Lancret and de Saint Venant~\cite{Struik1988},
I am unaware of a proof of it in the biological literature,
where it often is illustrated by photographs
of stacks of identical blocks, each rotated by a fixed
angle about the vertical axis~\cite{MBoC4151}\@.
Because of its far-reaching implications for biology,
a proof that is simple, direct, and self-contained
should be useful.
Such a proof is given in Sec.~\ref{intro}\@.
In Sec.~\ref{formulas}, some formulas about
helices are derived.
In Sec.~\ref{examples}, the theorem and these formulas
are illustrated by and applied to 
nucleic acids, protein secondary structures, 
proteins, protein folding, and viral capsids.
The theorem implies, in particular, that 
the \(\beta\)-strand~\cite{Pauling1953a,Pauling1953b}, 
which is
the second most common secondary structure in proteins,
is a helix,
and that icosahedral viral capsids
are made of helices.
The paper ends with remarks about
helices and evolution.
\section{Regularity Implies Helicity\label{intro}}
Suppose we have a collection
of identical objects,
which we label with the integers.
Suppose each object has both
a socket and a knob.
Suppose that every knob 
can fit snuggly into every
socket and that, once seated,
no further rotation of the
knob in the socket is possible.
We can set the knob of  
object 1 into the socket of
object 2.  Then we can put
the knob of object 2 into the
socket of object 3.
Next, we can put
the knob of object 3 into the
socket of object 4.
If we continue in this way,
then the chain of objects
will form a helix defined by
the first three objects.  
\par
To see why, we fix our
attention on a selected point,
the same for all the objects.  We might
choose
the top of each socket.
Let's call the selected point
on the \(i\)th object \(\mathbf{p}_i\)\@.
Let \( \mathbf{a} = \mathbf{p}_2 - \mathbf{p}_1\),
so \( \mathbf{p}_2 = \mathbf{a} + \mathbf{p}_1\)\@.
The knob of each object 
protrudes from its object
in a way that is 
arbitrary but the same for
all our objects.
So the vector 
\(\mathbf{b} = \mathbf{p}_3 - \mathbf{p}_2\)
has the same length as \( \mathbf{a} \)
and is related to it by a 
\(3 \times 3 \) rotation matrix \( R \),
\( \mathbf{b} = R \, \mathbf{a} \)\@.
The rotation \( R \) includes
any extra rotation of \( \phi \) radians
about the \( \mathbf{b} \) axis
that may occur in the rigid motion that
attaches object 2 to object 1\@. 
So \( \mathbf{p}_3 = \mathbf{b} + \mathbf{p}_2
= R \, \mathbf{a} + \mathbf{a} + \mathbf{p}_1\)\@.
What about 
\(\mathbf{c} = \mathbf{p}_4 - \mathbf{p}_3\)\@?
Well, the lengths of all the vectors
\( \mathbf{p}_{i+1} - \mathbf{p}_i \)
are the same, and they are all related
by rotations.  And since the objects
are all identical, the vector \(\mathbf{c} \)
must be related to \( \mathbf{b} \) by
the rotation \( R \) in the rotated frame ---
that is,
\(\mathbf{c} = R \, R \, R^{-1} \, \mathbf{b} 
= R \, \mathbf{b} \)\@.
So \(\mathbf{c} = R^2 \, \mathbf{a} \),
and thus the point \(\mathbf{p}_4\) is given by
\( 
\mathbf{p}_4 = \mathbf{c} + \mathbf{p}_3 
= R^2 \, \mathbf{a} + R \, \mathbf{a} 
+ \mathbf{a} + \mathbf{p}_1 
\)\@.
The general rule is
\beq
\mathbf{p}_{n+2} = \sum_{k=0}^n 
R^k \, \mathbf{a} + \mathbf{p}_1.
\label {pn=}
\eeq
\par
Every rotation matrix \(R\) has
one real eigenvector \( \mathbf{\hat n} \)
with an eigenvalue of unity 
\( 
R \,\mathbf{\hat n} = \mathbf{\hat n} 
\label{Rn=n}
\)\@.
The eigenvector \( \mathbf{\hat n} \) 
is the axis of the rotation.
The caret means that the axis \( \mathbf{\hat n} \)
is normalized, and we fix its sign by
requiring that 
\( \mathbf{a} \cdot \mathbf{\hat n} \ge 0 \)\@.
(The other two eigenvectors \( \mathbf{e}_\pm \)
are complex with
unimodular eigenvalues 
\(
R \, \mathbf{e}_\pm = e^{\pm i \theta} \,
\mathbf{e}_\pm 
\label{Re+-}
\)
in which \( \theta \) is the angle
of the rotation \( R \)\@.)
\par
Let us adopt a coordinate system
in which the \(z\) axis is
the axis of rotation
\( \mathbf{\hat z} = \mathbf{\hat n} \),
and the vector \( \mathbf{a} \) lies
in the \(x-z\) plane
\( \mathbf{a} = a_x \, \mathbf{\hat x}
+ a_z \, \mathbf{\hat z} \)\@.
The rotation now is about the \(z\) axis,
and so
\beq
R \, \mathbf{a} =
R \left( a_x \, \mathbf{\hat x}
+ a_z \, \mathbf{\hat z} \right)
= a_x \left( \cos \theta \, \mathbf{\hat x}
+ \sin \theta \, \mathbf{\hat y} \right)
+  a_z \, \mathbf{\hat z}
\label{Ra}
\eeq
in which we choose to have \( - \pi < \theta \le \pi \)\@.
If the product \( \theta \, a_z \) is positive,
then the helix is right handed;
if it is negative, then the helix is left handed.
The \(k\)th power of \( R \) turns \( \mathbf{a} \)
into
\beq
R^k \mathbf{a} =
R^k \left( a_x \, \mathbf{\hat x}
+ a_z \, \mathbf{\hat z} \right)
= a_x \left( \cos k\theta \, \mathbf{\hat x}
+ \sin k\theta \, \mathbf{\hat y} \right)
+  a_z \, \mathbf{\hat z}.
\label{Rka}
\eeq
So formula (\ref{pn=})
for the point \(\mathbf{p}_{n+2}\) gives 
\beq
\mathbf{p}_{n+2} =  (n+1) a_z \mathbf{\hat z} 
+ \mathbf{p}_1 
+ a_x \sum_{k=0}^n
 \left( \cos k\theta \, \mathbf{\hat x}
+ \sin k\theta \, \mathbf{\hat y} \right).
\label{pnis}
\eeq
Now by expressing \( \sin k\theta \)
and \( \cos k\theta \) in terms 
of \( \exp(i\theta) \) and by using
the relation
\(
(z-1) \sum_{k=0}^n z^k
= z^{n+1} - 1
\label{1 + z +}
\), 
one may derive the trigonometric
identities
\beq
\sum_{k=0}^n \cos k\theta = \thalf \left(
\cos n\theta \, +
\cot\thalf \theta \, \sin n \theta \, 
+ 1 \right)
\label{sumcos}
\eeq
and
\beq
\sum_{k=0}^n \sin k\theta = 
\thalf \left[
\sin n\theta \, -
\cot\thalf \theta \, ( \cos n \theta \, 
- 1) \right].
\label{sumsin}
\eeq
By substituting these identities
into Eq.(\ref{pnis}), we find
\bea
\mathbf{p}_{n+2} & = & (n+1) a_z \mathbf{\hat z} 
+ \mathbf{p}_1 \nn\\
& + & \thalf a_x 
\left(
\cos n\theta \, +
\cot\thalf \theta \, \sin n \theta \, 
+ 1 \right) \mathbf{\hat x} \nn\\
& + & \thalf a_x 
\left[
\sin n\theta \, -
\cot\thalf \theta \, ( \cos n \theta \, 
- 1) \right]  \mathbf{\hat y} .
\label{pnisr}
\eea
If we call \( \mathbf{v} \)
the vector
\beq
\mathbf{v} = \frac{a_x}{2} \left( \begin{array} {c}
1 \\ - \cot \thalf \theta \\ 0
\end{array} \right)
\eeq
then we may write the general point
\( \mathbf{p}_{n+2} \) as
\beq
\mathbf{p}_{n+2} = R^n \mathbf{v} + n a_z \mathbf{\hat z} 
+ \mathbf{p}_1 + \mathbf{a} - \mathbf{v}
\label{pnisf}
\eeq
which clearly is a helix.
\par
A rotation \( R \) about a point \( \mathbf{p_0} \)
takes a point \( \mathbf{p} \) into the point
\( \mathbf{p}' \) given by
\beq
\mathbf{p}' -  \mathbf{p}_0 = R 
\left( \mathbf{p} -  \mathbf{p}_0 \right).
\label{how world turns}
\eeq
By comparing this rule with our formula (\ref{pnisf})
for \( \mathbf{p}_{n+2} \), we may infer that
\( \mathbf{v} = a_x \hat{\mathbf{x}} - \mathbf{r}_0 \)
in which \( \mathbf{r}_0 \) is
the point where 
the axis \( \hat{\mathbf{z}} \) crosses
the \(x-y\) plane; equivalently
\beq
\mathbf{r}_0 = a_x \hat{\mathbf{x}} - \mathbf{v}
= \thalf \, a_x \,
\left[ \, \hat{\mathbf{x}} + 
\cot(\theta/2) \, \hat{\mathbf{y}} \,
\right].
\label{pointOnAxis}
\eeq
Equation (\ref{pnisf}) for 
the point \( \mathbf{p}_{n+2} \) now takes the form
\beq
\mathbf{p}_{n+2} - \mathbf{r}_0 =
R^n ( a_x \hat{\mathbf{x}} - \mathbf{r}_0 )
+ (n+1) a_z \hat{\mathbf{z}} + \mathbf{p}_1
\label{pnNice}
\eeq
or more simply
\beq
\mathbf{p}_{n+2} - \mathbf{r}_0 =
R^n ( {\mathbf{a}} - \mathbf{r}_0 )
+ n \, a_z \hat{\mathbf{z}} + \mathbf{p}_1
\label{pnSimply}
\eeq
since \( R^n \hat{\mathbf{z}} =  \hat{\mathbf{z}} \)\@.
\par
This helix rises by \( \Delta z = a_z \)
with each object and turns by the
angle \( \theta \) with each object,
so its pitch is 
\(p = (2\pi/\theta) \, \Delta z = 2 \pi a_z / \theta\)\@.
Its axis is \( \mathbf{r}_0 + z \, \hat{\mathbf{z}}\)
for all \(z\)\@.
\par
The rotation matrix \({R}\) is
the product of a rotation \( R(\mathbf{b},\mathbf{a}) \)
that rotates the vector \( \mathbf{a} \) into
the vector \( \mathbf{b} \)\@ and a rotation 
\( R(\phi \hat{\mathbf{b}}) \) about
the vector \( \mathbf{b} \) by a 
dihedral angle \( \phi \)
\beq
R = R(\phi \hat{\mathbf{b}}) \, R(\mathbf{b},\mathbf{a}).
\label{R=RR}
\eeq
The first matrix \( R(\mathbf{b},\mathbf{a}) \) is
\bea
R(\mathbf{b},\mathbf{a}) & = &
| \widehat{\mathbf{a} \times \mathbf{b}} \rangle 
\langle  \widehat{\mathbf{a} \times \mathbf{b}} |
+
| \hat{\mathbf{b}} \rangle \langle \hat{\mathbf{a}} | \nn\\
& + &
| \widehat{(\mathbf{a} \times \mathbf{b}) \times \mathbf{b} } \rangle 
\langle  \widehat{(\mathbf{a} \times \mathbf{b}) \times \mathbf{a} } |
\label{Rba}
\eea
in Dirac notation with the carets meaning
that all the vectors are unit vectors.
The second matrix \( R(\phi \hat{\mathbf{b}}) \) is~\cite{Cahill2000}
\beq
R(\phi \hat{\mathbf{b}}) =
e^{\phi \hat{\mathbf{b}} \cdot \vec L} = 
\cos \phi \, I + \hat \mathbf{b} \cdot \vec L \, \sin \phi
+ ( 1 - \cos \phi ) \, \hat \mathbf{b} ( \hat \mathbf{b} )^\mathsf{T}
\eeq
in which the generators \( (L_k)_{ij} = \epsilon_{ikj} \) satisfy
\( [ L_i, L_j ] = \epsilon_{ijk} L_k \)
and \( \mathsf{T} \) means transpose.
In terms of indices, this formula for
\( R( \phi \hat{\mathbf{b}} ) 
= e^{\phi \hat{\mathbf{b}} \cdot \vec L} \) 
is
\beq
R( \phi \hat{\mathbf{b}} )_{ij} =
\delta_{ij} \cos \phi - \sin \phi \, \epsilon_{ijk} \hat \mathbf{b}_k
+ ( 1 - \cos \phi ) \, \hat \mathbf{b}_i \hat \mathbf{b}_j.
\eeq
In these formulas, \( \epsilon_{ijk} \)
is totally antisymmetric with \( \epsilon_{123}=1 \),
and sums over \(k\) from 1 to 3 are understood.

\section{Parametrizing a Helix\label{formulas}}
Suppose you are given a set of points
that lie on a helix.  How do you
find the spacing \( \Delta z \), 
the angle \( \theta \) per step,
the axis \( \mathbf{\hat n} \),
and a point \( \mathbf{n}_0 \) on the helix?
A helix is defined by four points 
\(\mathbf{p}_1\), \(\mathbf{p}_2\), 
\(\mathbf{p}_3\), \(\mathbf{p}_4\).
Let \(\mathbf{a} = \mathbf{p}_2 - \mathbf{p}_1\),
\(\mathbf{b} = \mathbf{p}_3 - \mathbf{p}_2\), and
\(\mathbf{c} = \mathbf{p}_4 - \mathbf{p}_3\)\@.
If the axis of the helix points in the
direction \(\mathbf{\hat n}\) 
and  \(\mathbf{n}_0\) is any point
on the axis, then 
the axis contains the
points \(\mathbf{n}_0 + z \, \mathbf{\hat n}\),
where \(z\) is any real number.
\par
The points \(\mathbf{p}_i\) of the helix
are evenly spaced by \(\Delta z\)
in the \(\mathbf{\hat n}\) direction.
The spacing \( \Delta z\) is given
by
\beq
\Delta z = \mathbf{\hat n} \cdot 
(\mathbf{p}_2 - \mathbf{p}_1)
= \mathbf{\hat n} \cdot \mathbf{a}.
\label{n.a}
\eeq
Because it is constant, the spacing 
\(\Delta z\) is also given by
\( 
\Delta z = \mathbf{\hat n} \cdot 
(\mathbf{p}_3 - \mathbf{p}_2)
= \mathbf{\hat n} \cdot \mathbf{b}
\label{n.b}
\) 
and by
\( 
\Delta z = \mathbf{\hat n} \cdot 
(\mathbf{p}_4 - \mathbf{p}_3)
= \mathbf{\hat n} \cdot \mathbf{c}.
\label{n.c}
\) 
Thus the axis \(\mathbf{\hat n}\) is
orthogonal to \(\mathbf{b} -\mathbf{a}\)
and to \(\mathbf{c} -\mathbf{b}\)\@.
So it must be
parallel (or antiparallel)
to the cross product \(\mathbf{n}\)
of these vectors,
\beq
\mathbf{n} = ( \mathbf{b} -\mathbf{a} ) \times
( \mathbf{c} -\mathbf{b} ).
\label{ndir}
\eeq
In terms of the length 
\(\ell = |\mathbf{b} -\mathbf{a} | = 
| \mathbf{c} -\mathbf{b} | \)
and the angle \(\phi\)
between the vectors \(\mathbf{b} -\mathbf{a} \)
and \(\mathbf{c} -\mathbf{b} \),
the vector \(\mathbf{n}\) is of length
\(\ell^2 \sin \phi\)\@.
The general direction of the helix
is defined by the difference
\(\mathbf{p}_4 - \mathbf{p}_1 = 
\mathbf{a} + \mathbf{b} + \mathbf{c} \);
so if \(\sigma\) is the sign
of the dot product 
\((\mathbf{a} + \mathbf{b} + \mathbf{c} )
\cdot \mathbf{n}\),
then the axis of the helix is
the unit vector
\beq
\mathbf{\hat n} = \sigma \, 
\frac{ ( \mathbf{b} -\mathbf{a} ) \times
( \mathbf{c} -\mathbf{b} )}{ \ell^2 \, \sin \phi}.
\label{nhat}
\eeq
\par
The three other parameters of the helix
are its radius \(\rho\), its angle
\(\theta\), and
a point \(\mathbf{n}_0\)
on its axis.
To find these, we note that
for each of the four points
\(\mathbf{p}_i\),
\beq
\left[ \, \mathbf{\hat n} \times 
(\mathbf{p}_i - \mathbf{n}_0) \, \right]^2
= \rho^2.
\label{norho}
\eeq
Subtracting this relation for \(i = 1\)
from this relation for \(i = 2\)
and recalling that \(\mathbf{a} = p_2 -p_1\),
we get an equation that is linear in 
\(\mathbf{n}_0\)
\beq
2 \, ( \mathbf{\hat n}  \times \mathbf{a} ) \cdot 
( \mathbf{\hat n}  \times \mathbf{n}_0 ) = 
( \mathbf{\hat n}  \times \mathbf{p}_2 )^2 
- ( \mathbf{\hat n}  \times \mathbf{p}_1 )^2.
\label{quad21}
\eeq
Similarly, subtracting Eq.(\ref{norho})
for \(i = 2\) from Eq.(\ref{norho})
for \(i = 3\) and recalling that 
\(\mathbf{b} = \mathbf{p}_3 - \mathbf{p}_2\),
we find
\beq
2 \, ( \mathbf{\hat n}  \times \mathbf{b} ) \cdot 
( \mathbf{\hat n}  \times \mathbf{n}_0 ) = 
( \mathbf{\hat n}  \times \mathbf{p}_3 )^2 
- ( \mathbf{\hat n}  \times \mathbf{p}_2 )^2.
\label{quad32}
\eeq
\par
An orthonormal basis is provided by
the three vectors
\beq
\mathbf{\hat e}_1  =  
\frac{ \mathbf{b} - \mathbf{a} }{|\mathbf{b} - \mathbf{a}|}, \quad
\mathbf{\hat e}_2  =  \mathbf{\hat n} \times \mathbf{\hat e}_1, \quad
\mathrm{and} \quad 
\mathbf{\hat e}_3  =  \mathbf{\hat n} .
\label{e}
\eeq
Subtracting Eq.(\ref{quad21}) from Eq.(\ref{quad32}),
we get
\bea
\lefteqn{2 \, \left[ \mathbf{\hat n} \times 
(\mathbf{b} - \mathbf{a} ) \right]
\cdot ( \mathbf{\hat n}  \times \mathbf{n}_0 ) = \quad} \nn\\ 
& & ( \mathbf{\hat n}  \times \mathbf{p}_3 )^2 
- 2 \, ( \mathbf{\hat n}  \times \mathbf{p}_2 )^2
+ ( \mathbf{\hat n}  \times \mathbf{p}_1 )^2 
\label{n1}
\eea
or, using (\ref{e}),
\bea
\lefteqn{2 \, | \mathbf{b} - \mathbf{a} | \,
( \mathbf{\hat n} \times \mathbf{\hat e}_1 ) \cdot
( \mathbf{\hat n}  \times \mathbf{n}_0 ) = } \nn\\
& & ( \mathbf{\hat n}  \times \mathbf{p}_3 )^2 
- 2 \, ( \mathbf{\hat n}  \times \mathbf{p}_2 )^2
+ ( \mathbf{\hat n}  \times \mathbf{p}_1 )^2.
\label{n2}
\eea
So in terms of the definition
\beq
C_{321} = \frac{( \mathbf{\hat n}  \times \mathbf{p}_3 )^2 
- 2 \, ( \mathbf{\hat n}  \times \mathbf{p}_2 )^2
+ ( \mathbf{\hat n}  \times \mathbf{p}_1 )^2}
{2 \, | \mathbf{b} - \mathbf{a} |},
\label{C321}
\eeq
we may use (\ref{e}) again to write Eq.(\ref{n2}) as
\beq
\mathbf{\hat e}_2 \cdot
( \mathbf{\hat n}  \times \mathbf{n}_0 ) = C_{321}.
\label{n3}
\eeq
\par
Since the unit vectors \(\mathbf{\hat e}_i\)
are complete and orthonormal,
we may expand the axis point 
\(\mathbf{n}_0\) as
\beq
\mathbf{n}_0 = \sum_{i=1}^3
(\mathbf{\hat e}_i \cdot \mathbf{n}_0) \, \mathbf{\hat e}_i .
\label{n0}
\eeq 
Using (\ref{e}) and the relations
\( \mathbf{\hat n} \times \mathbf{\hat e}_2 = - \mathbf{\hat e}_1\)
and \(  \mathbf{\hat n} \times \mathbf{\hat e}_3 
= \mathbf{0} \), 
we have
\beq
 \mathbf{\hat n}  \times \mathbf{n}_0 =
(\mathbf{\hat e}_1 \cdot \mathbf{n}_0) \, \mathbf{\hat e}_2 
- (\mathbf{\hat e}_2 \cdot \mathbf{n}_0) \, \mathbf{\hat e}_1.
\label{nxn0}
\eeq
So Eq.(\ref{n3}) now implies
\beq
\mathbf{\hat e}_1 \cdot \mathbf{n}_0 = C_{321}.
\label{e1n0}
\eeq
\par
Expanding the vector \(\mathbf{a}\) 
in terms of the basis \(\{\mathbf{\hat e}_i\}\)
\beq
\mathbf{a} = \sum_{i=1}^3
(\mathbf{\hat e}_i \cdot \mathbf{a}) \, \mathbf{\hat e}_i,
\label{a}
\eeq 
and using (\ref{e}), we find
\beq
\mathbf{\hat n} \times \mathbf{a} = 
(\mathbf{\hat e}_1 \cdot \mathbf{a}) \, \mathbf{\hat e}_2
- (\mathbf{\hat e}_2 \cdot \mathbf{a}) \, \mathbf{\hat e}_1.
\label{nxa}
\eeq
This relation and Eq.(\ref{nxn0}) imply
\beq
( \mathbf{\hat n} \times \mathbf{a} ) \cdot
(\mathbf{\hat n} \times \mathbf{n}_0 ) =
(\mathbf{\hat e}_1 \cdot \mathbf{a}) \,
(\mathbf{\hat e}_1 \cdot \mathbf{n}_0) +
(\mathbf{\hat e}_2 \cdot \mathbf{a}) \,
(\mathbf{\hat e}_2 \cdot \mathbf{n}_0).
\label{nxa.nxn0}
\eeq
Using this expression and the notation
\beq
C_{21} = \frac{1}{2} \, \left[
( \mathbf{\hat n}  \times \mathbf{p}_2 )^2 
- ( \mathbf{\hat n} \times \mathbf{p}_1 )^2
\right],
\label{C21}
\eeq
we extract from Eq.(\ref{quad21}) the result
\beq
(\mathbf{\hat e}_1 \cdot \mathbf{a}) \,
(\mathbf{\hat e}_1 \cdot \mathbf{n}_0) +
(\mathbf{\hat e}_2 \cdot \mathbf{a}) \,
(\mathbf{\hat e}_2 \cdot \mathbf{n}_0) =
C_{21}
\eeq
or, using (\ref{e1n0}),
\beq
\mathbf{\hat e}_2 \cdot \mathbf{n}_0 = \frac
{C_{21} - (\mathbf{\hat e}_1 \cdot \mathbf{a}) \, C_{321}}
{\mathbf{\hat e}_2 \cdot \mathbf{a}}.
\label{e2n0}
\eeq
The inner product 
\( \mathbf{\hat e}_3 \cdot \mathbf{\hat n}_0 \)
is arbitrary.  So by substituting our formulas
(\ref{e1n0}) for 
\( \mathbf{\hat e}_1 \cdot \mathbf{n}_0 \)
and (\ref{e2n0}) for 
\( \mathbf{\hat e}_2 \cdot \mathbf{n}_0 \)
into the expansion (\ref{n0}), we have a set
of points \( \mathbf{n}_0 \) on the axis
of the helix in terms of the free parameter
\( \mathbf{\hat e}_3 \cdot \mathbf{\hat n}_0 \)\@.
\par
To find the radius \(\rho\) from Eq.(\ref{norho}),
we use Eq.(\ref{nxn0}) for 
the cross product 
\(\mathbf{\hat n} \times \mathbf{n}_0\):
\beq
\rho = \sqrt{
\left[ \mathbf{\hat n} \times \mathbf{p}_1
+ ( \mathbf{\hat e}_2 \cdot 
\mathbf{n}_0 ) \, \mathbf{\hat e}_1
- ( \mathbf{\hat e}_1 \cdot 
\mathbf{n}_0 ) \, \mathbf{\hat e}_2
\right]^2
},
\label{rho}
\eeq
where the axis \(\mathbf{\hat n}\)
and the inner products
\(\mathbf{\hat e}_2 \cdot \mathbf{n}_0\)
and
\(\mathbf{\hat e}_1 \cdot \mathbf{n}_0\)
are given respectively by 
(\ref{nhat}), (\ref{e1n0}), and (\ref{e2n0})\@.
\par
The cosine of the angle \(\theta\) is
\beq
\cos \theta =
\rho^{-2} \,
[\mathbf{\hat n} \times 
( \mathbf{p}_1 - \mathbf{n}_0 )]
\cdot
[\mathbf{\hat n} \times 
( \mathbf{p}_2 - \mathbf{n}_0 )],
\label{cosDphi}
\eeq
and its sine is
\beq
\sin \theta =
\rho^{-2} \,
\mathbf{\hat n} \cdot
[\mathbf{\hat n} \times 
( \mathbf{p}_1 - \mathbf{n}_0 )]
\times
[\mathbf{\hat n} \times 
( \mathbf{p}_2 - \mathbf{n}_0 )].
\label{sinDphi}
\eeq
So the angle \(\theta\) is
the argument of the complex number
\((\cos \theta, \sin \theta)\)
in the interval \( -\pi < \theta \le \pi \),
which is given by the \textsc{fortran}
arctangent function \texttt{atan2} as
\(
\theta = \mathtt{atan2}
(\sin \theta, \cos \theta)
\label{Dphi}
\)\@.
\section{Examples of Bio-Helices\label{examples}}
\textbf{DNA:}
Although DNA is made out of 
nucleotides, its building block
is the object dR-B-B\(^\prime\)-dR
in which 
dR is a deoxyribose sugar and 
B-B\(^\prime\) is a Crick-Watson
base pair of adenine and thymine 
(A=T or T=A) or of cytosine and guanine
(C\(\equiv\)G or G\(\equiv\)C)\@.   
The four base pairs
have nearly the same size, and so the four units
dR-B-B\(^\prime\)-dR are nearly identical.
Phosphate groups glue these units into 
a regular chain.  Each dR is linked by one 
phosphate group to the unit behind it
and by another phosphate group to 
the unit ahead of it.  This pattern
of covalent bonds is nearly the same in all 
dR-B-B\(^\prime\)-dR units.
The result is a helix or
a double helix~\cite{Watson1953} if one
counts both chains of sugar-phosphate groups.
For instance,
the ideal B-DNA dodecamer d(CGCGAATTCGCG)
at 1.4 \AA\ resolution 
is a right-handed double helix with
\( a_z = \Delta z = 3.3 \) \AA,
\( \theta =  35.5^\circ \), a
diameter of 20 \AA, and a 
pitch of 33.3 \AA~\cite{Williams2000}\@.
But other sequences of base pairs
have \( \theta \) as low as
\(26^\circ \) or as high as \( 43^\circ \)\@.
When the relative humidity is below 75\%,
B-DNA turns into the  
A form, which 
is a right-handed helix with 
a pitch of 34 \AA, but with
\( \theta = 26^\circ \)\@.
The Z form, which can occur when 
the salt concentration is high, 
is a left-handed helix with 
\( \theta = 18^\circ \) and a pitch
of 44 \AA\@.
The dodecamer d(\((AT)^6\)) forms 
coiled coils~\cite{Campos2005}\@.
\par
\textbf{Secondary Structures of Proteins:}
Proteins are chains of amino acids.
Except for proline, the 
20 amino acids
differ only in their side chains.
The amino acids all have 
the same main chain N\(-\)C\(-\)C and 
are linked together 
N\(-\)C\(-\)C\(\frac{\dots}{}\)N\(-\)C\(-\)C
\(\frac{\dots}{}\)N\(-\)C\(-\)C
by peptide bonds,
which resist rotations --- the angle \( \omega \)
about the C\(\frac{\dots}{}\)N 
bond usually is close to \( 180^\circ \)\@.
The dihedral angles \( \phi \) and \( \psi \)
describe rotations about the 
axes of the single bonds N\(-\)C and C\(-\)C\@.
These angles
are the principal degrees of freedom
in proteins, but
they are far from free.
Ramachandran steric constraints
force them to lie in three regions,
more (proline) or less (glycine)\@.
\par
The \( \alpha \) region lies near 
\( \phi = - 57^\circ \), \( \psi = - 47^\circ \),
and \( \omega = 180^\circ \)~\cite{Dover1967}\@.
A chain of amino acids with these dihedral angles
is an \( \alpha \)-helix~\cite{Pauling1951a}\@.
By using Eqs.~(\ref{n.a}--\ref{sinDphi}),
one may show that the ideal \( \alpha \)-helix
is right handed and that it
has 3.62 residues per turn,
\( \theta = 99.4^\circ \), \( a_z = \Delta z = 1.56 \) \AA,
and a pitch of 5.64 \AA\@.
This geometry allows the carboxyl oxygen of the \(i\)th
amino acid to flirt with the hydrogen
of the main-chain nitrogen of the \(i+4\)th 
amino acid; the energy of the resulting
N\(_{i+4}\)\(-\)H\(\!\,\,\textstyle{\cdot\!\cdot\!\cdot}\,\!\,\)O\(=\)C\(_{i}\)
hydrogen bond is of the order of 0.3 eV\@.
\par
The other two sterically allowed regions 
are side by side.
The more important one, near
\( \phi = - 139^\circ \), \( \psi = 135^\circ \),
and \( \omega = -178^\circ \)~\cite{Arnott1967},
generates helices that form hydrogen bonds
between their main-chain amino
and carboxyl groups when the helices are 
adjacent and antiparallel, forming 
an antiparallel \( \beta \)-sheet.
Formulas (\ref{n.a}--\ref{sinDphi}) imply that
the ideal antiparallel \( \beta \)-helix
has 2.004 residues per turn,
\( a_z = \Delta z = 3.47 \) \AA, and
a pitch of 6.95 \AA\@.
Although slightly left handed
with \( \theta = - 179.7^\circ \),
it is nearly planar.
Changes of \( \phi \), \( \psi \), and
\( \omega \) by \( 1^\circ \) flip
the angle \( \theta \) 
across the cut at \( \theta = \pi \); so
antiparallel \( \beta \)-helices
do not have definite helicity.
\par
The other region, near
\( \phi = - 119^\circ \), \( \psi = 113^\circ \),
and \( \omega = 180^\circ \)~\cite{Schellman1964},
generates helices that form hydrogen bonds
between their main-chain amino
and carboxyl groups when the helices are 
adjacent and parallel --- a parallel \( \beta \)-sheet.
The ideal parallel \( \beta \)-helix
has 2.024 residues per turn,
\( a_z = \Delta z = 3.27 \) \AA, and
a pitch of 6.62 \AA\@.
Although somewhat right handed
with \( \theta = 177.8^\circ \),
it is nearly planar.
Changes of \( \phi \), \( \psi \), and
\( \omega \) by 3 or \( 4^\circ \) can
flip the angle \( \theta \) 
across the cut at \( \theta = \pi \), and so
parallel \( \beta \)-helices
do not have definite helicity.
In a parallel \( \beta \)-sheet,
the distance along its main chain 
between an amino group and
the carboxyl group to which it hydrogen-bonds
is greater than in an antiparallel
\( \beta \)-sheet (or an \( \alpha \)-helix), and so
proteins with parallel \( \beta \)-sheets 
fold slowly.
\par
Students would get a more unified
view of secondary structure in proteins and nucleic acids 
if authors of biochemistry textbooks
called \( \beta \)-strands ``\( \beta \)-helices.''
\par
\textbf{Proteins:}
The main fibrous protein in
hair, horn, and nails,
\( \alpha \)-keratin,
is two 
\( \alpha \)-helices wrapped
around each other in a left-handed double helix.
The key protein of
the extracellular matrix holding cells
in animal tissue, collagen,
is three 
\( \alpha \)-helices 
in a right-handed triple helix.
Actin and tubulin form helical cytoskeletal filaments.
\par
Globular and transmembrane proteins are 
\( \alpha \)- and \( \beta \)-helices linked by
loops and turns.
They can be as dense as crystals;
\( \alpha \)-helices pack
closely~\cite{Banavar2000}\@.
\par
\textbf{A Remark about Protein Folding:}
Proline aside, any string of amino acids can fold
into an \( \alpha \) or 
a \( \beta\)-helix.  How does it decide? 
The solvent helps it decide.
Proteins fold in salty water, a polar solvent.
A protein in a polar solvent
has a lower energy if 
the hydrophilic (charged or polar)
side chains are on the
outside and the hydrophobic
ones are inside.  
Suppose
two hydrophilic side chains are
separated on the main chain by \(n\)
hydrophobic ones.
They will form a helix
that cuts through 
the protein on a chord 
of length \( n \, \Delta z \) with the
two hydrophilic ones outside
at its ends.
But \( \Delta z_\beta = 3.4\)~\AA\ for a \(\beta\)-helix
is twice \( \Delta z_\alpha = 1.6\)~\AA\
for an \(\alpha\)-helix.
So the choice between the two kinds
of helices is decided in part by whether 
\( n \, \Delta z_\alpha \) or
\( n \, \Delta z_\beta \) 
is closer to the thickness of the protein.
\par
\textbf{Viral Capsids}
The coats (or capsids) of
filamentary viruses often are
made of a single helix.  For instance,
the helical capsid of the tobacco mosaic virus consists
of 2130 copies of a single protein~\cite{Caspar1998}\@.
\par
The coats of icosahedral viruses~\cite{Baker1999} 
are made of nested helices in which
\(T = h^2 + hk + k^2 = 1\), 3, 4, 7, \(\dots\) 
protein molecules form 
an ``asymmetric unit.''
Three of these asymmetric units form a triangular,
primary helix of zero pitch.
In turn, ten of these primary, triangular helices
form two pentagonal, zero-pitch, secondary helices,
and another ten of them
form a secondary, beltlike, zero-pitch deca-helix.
An icosahedron results when 
the two secondary pentagonal helices attach to
opposite sides of the secondary belt of ten
triangular helices.
The capsid is made of \(60 T\) protein molecules.
\par
\textbf{Why all the helices?}
Evolution, size, and geometry require them.
Evolution forces economy. 
Cells use mass production
to achieve economy, 
making many copies of identical or 
closely related objects.
Cells are too small to have 
workers, so
they use macromolecules.
DNA polymerase 
makes DNA; RNA polymerase 
makes RNA; ribosomes make proteins;
proteins fold 
automatically or via chaperones;
protein complexes 
self-assemble.
These processes 
assemble nearly identical objects 
in regular ways.
Helices are ubiquitous because
identical objects, regularly assembled, 
form helices.
\begin{acknowledgments}
Thanks to S.~Atlas, E.~Coutsias, D.~Cromer,
K.~Dill, T.~Duke, A.~Parsegian, R.~Radloff, 
D.~Sergatskov, J.~Thomas,
and T.~Tolley
for advice.
\end{acknowledgments}
\bibliography{chem,cs,physics,vdw,bio,proteins,math,biochem}

\end{document}